\documentclass[12pt]{article}

\usepackage{graphics,graphicx}
\usepackage{aastexdefs,amssymb}
\usepackage{times}
\makeatletter
\renewcommand{\section}{\@startsection%
{section}{1}{0mm}{-\baselineskip}%
{0.5\baselineskip}{\normalfont\Large\bfseries}}%
\makeatother

\begin{document}

\thispagestyle{empty}

\begin{center}

{\bf {\large The Evolution of Galaxy Clusters Across Cosmic Time}}

\vspace{0.5in}

{\bf {\large A Science Working Paper for the 2010 Decadal Survey}}

\vspace{1in}

M.~Arnaud$^{1}$,
H.~Bohringer (MPE), C.~Jones (CfA), B.~McNamara (University of 
Waterloo), T.~Ohashi (Tokyo Metropolitan University),
D.~Patnaude (CfA), K.~Arnaud (NASA/GSFC), M.~Bautz (MIT), 
A.~Blanchard (Laboratorie d'Astrophysique de Toulouse-Tarbes), 
J.~Bregman (University of Michigan), G.~Chartas (Penn State University),
J.~Croston (University of Hertfordshire), L.~David (CfA), 
M.~Donahue (Michigan State University), A.~Fabian (IfA, Cambridge), 
A.~Finoguenov (MPE), A.Furuzawa (Nagoya University),
S.~Gallagher (University of Western Ontario), Y.~Haba (Nagoya University), 
A.~Hornschemeier (NASA/GSFC), S.~Heinz (University of Wisconsin), 
J.~Kaastra (SRON), W.~Kapferer (University of Innsbruck), 
G.~Lamer (Astrophysikalisches Institut Potsdam), 
A.~Mahdavi (San Francisco State University), 
K.~Makishima (The University of Tokyo),
K.~Matsushita (Tokyo University of Science), 
K.~Nakazawa (The University of Tokyo), 
P.~Nulsen (CfA),  P.~Ogle (Spitzer Science Center), 
E.~Perlman (University of Maryland, Baltimore County), 
T.~Ponman (University of Birmingham), D.~Proga (University of Nevada, Las 
Vegas), G.~Pratt (MPE),
S.~Randall (CfA), T.~Reiprich (Argelander-Institut f\"{u}r Astronomie), 
G.~Richards (Drexel University),  K.~Romer (University of Sussex), 
M.~Ruszkowski (University of Michigan), 
R.~Schmidt (Astronomisches Rechen-Institut), 
R.~Smith (CfA),  H.~Tananbaum (CfA), A.~Vikhlinin (CfA), 
J.~Vrtilek (CfA),
D.~Worrall (University of Bristol)

\end{center}

\vfill
\noindent
$^{1}$CEA-IRFU \\
Service d'Astrophysique \\
Ormes des Merisiers \\
91191 Gif sur Yvette Cedex \\
France \\
email: Monique.Arnaud@cea.fr

\clearpage

\setcounter{page}{1}

The large scale structure of the present Universe is determined by the growth
of dark matter density fluctuations and by the dynamical action of dark energy
and dark matter. While much progress has been made in recent years in 
constraining the cosmological parameters, and in reconstructing the evolution
in the large--scale structure of the dark matter distribution, {\bf we still
lack an understanding of the evolution of the baryonic component of the 
Universe.}
How normal, baryonic matter collects in dark matter gravitational
wells and forms galaxies and clusters is not understood well enough
to make precise predictions about what we see, from first principles.

Present observations, as well as theoretical work, indicate that baryonic 
structure formation on various scales is deeply interconnected: galaxy
formation depends on the large scale environment in which galaxies are 
found, and on the physical and chemical properties of the intergalactic 
gas from which they form, which in turn is affected by galaxy feedback. Due
to the complex behavior of the baryonic matter, progress has been driven
largely by 
observations and requires us to study simultaneously the evolution
of the hot and cold components of the Universe. \\

Located at nodes of the cosmic web, clusters of galaxies are the largest
collapsed structures in the Universe with total masses up to 
10$^{15}$ M$_{\sun}$. Over 80\% of their mass resides in the form of dark
matter. The remaining mass is composed of baryons, most of which 
(about 85\%) is a diffuse, hot T $>$ 10$^{7}$ K plasma
(the intracluster medium, ICM) that radiates primarily in the 
X-ray band. Thus in galaxy clusters, through the radiation from the hot
gas and the galaxies, we can observe and study the interplay between the hot
and cold components of the baryonic matter and the dark matter. X-ray
observations of the evolving cluster population provide a unique
opportunity to address such open and fundamental questions as:
\begin{itemize}
\item How do hot diffuse baryons dynamically evolve in dark matter
potentials?
\item How and when was the excess energy which we observe in the intergalactic
medium generated?
\item What is the cosmic history of heavy-element production and
circulation?
\end{itemize}

Our current knowledge comes primarily from detailed studies of clusters in
the relatively nearby Universe (z$<$0.5). {\bf Major advances will come from 
high throughput, high spectral and spatial resolution X-ray observations that 
measure the thermodynamic properties and metal content of the first
low mass clusters emerging at z $\sim$ 2 and directly trace their
evolution into today's massive clusters.} X-ray observations at high spectral
resolution also will open completely new vistas in discovery space by directly
probing the dynamics of the hot gas by mapping the velocity field and
turbulence.


\begin{figure*}[tbh!]
\begin{minipage}[c]{1.0\textwidth}
\includegraphics[bb=87 406 524 723,scale=0.55]{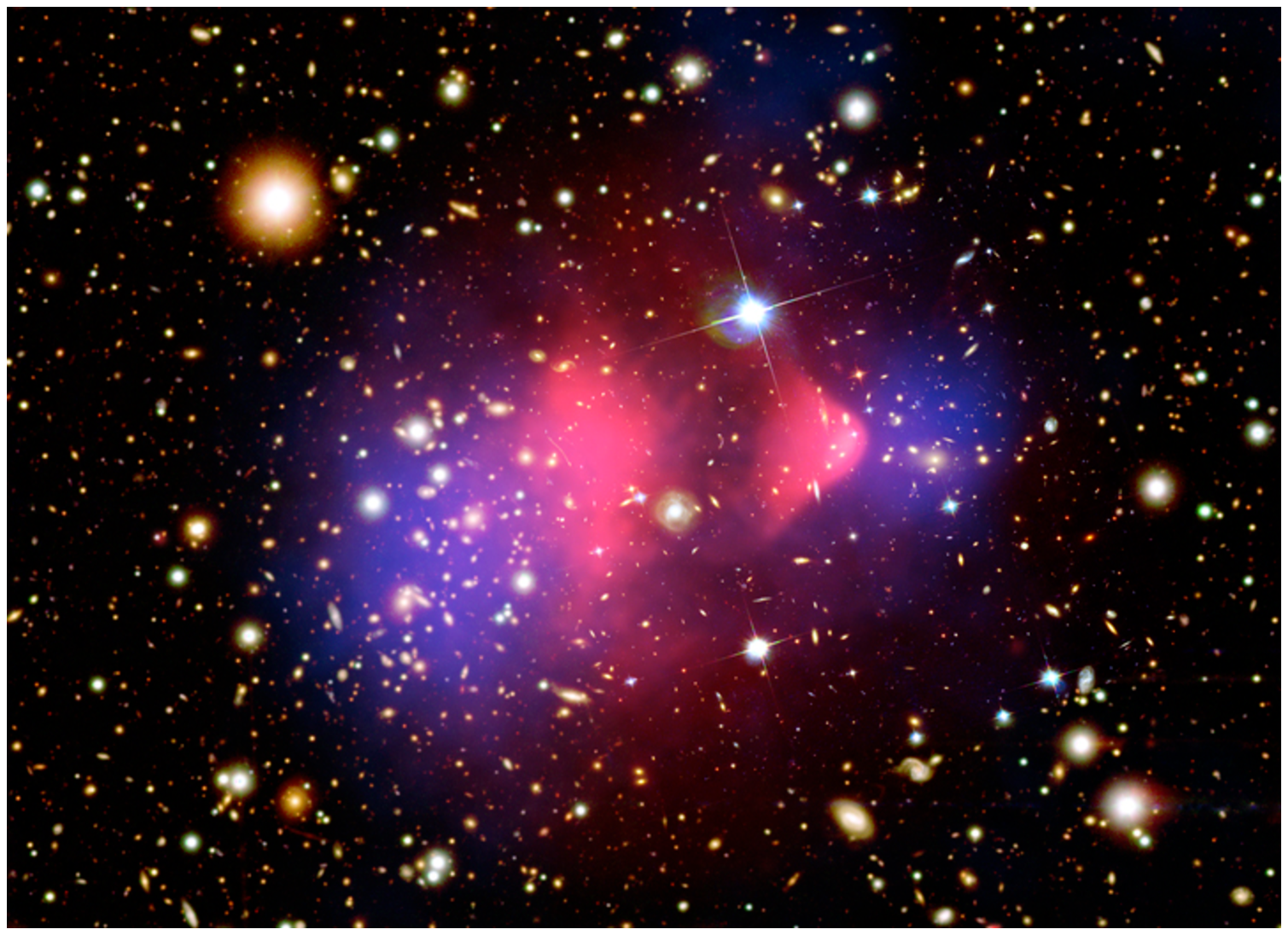}
\hfill \includegraphics[bb=89 268 525 722,scale=0.4]{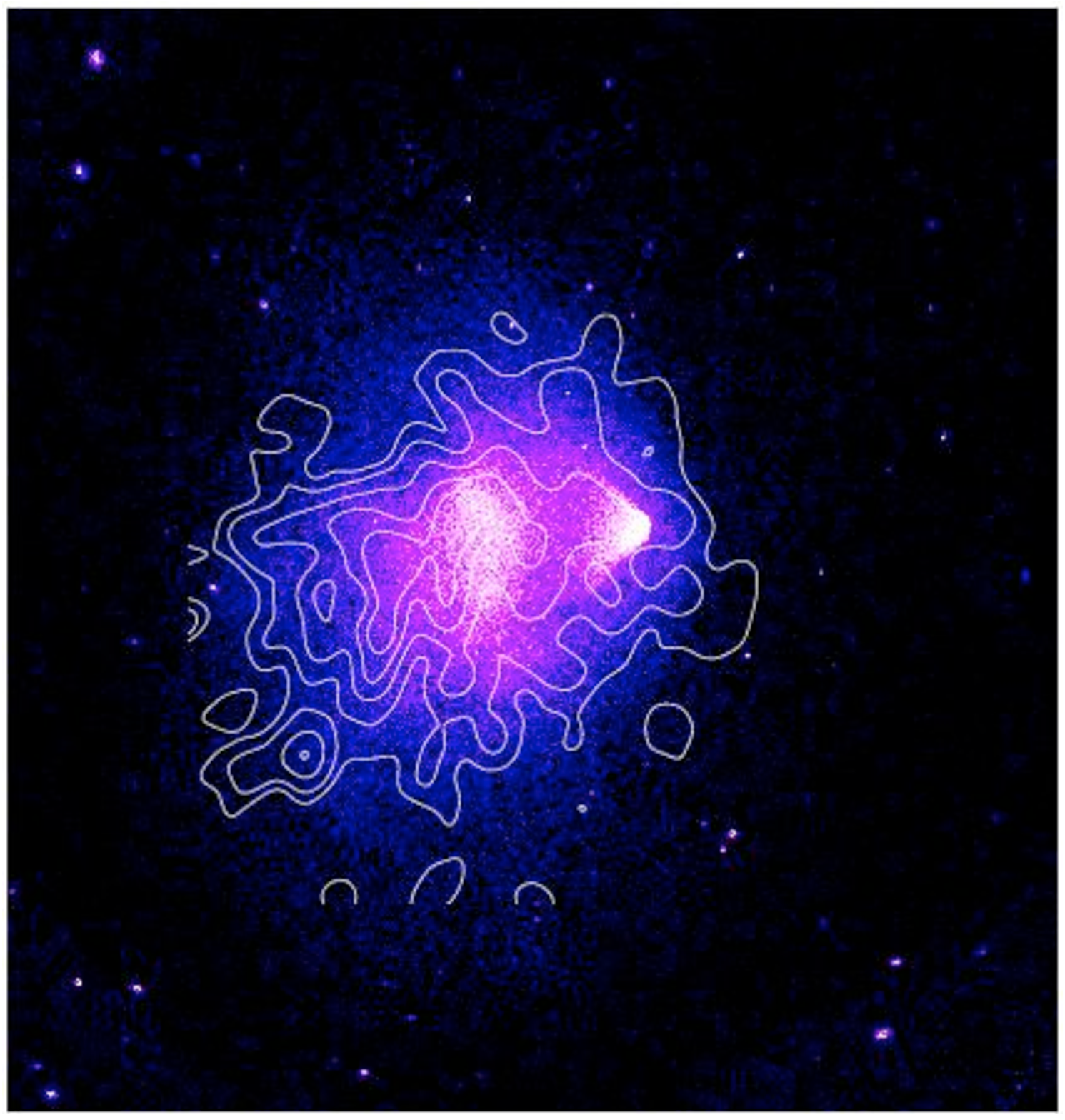}
\end{minipage}
\caption{{\bf Left:} As shown here for the ``Bullet Cluster'' 
(1E 0657-56), following the subcluster -- cluster collision,
the X-ray gas (in pink) and the dark matter as traced by the 
lensing (in blue) can become separated. {\bf Right:} Contours of 
radio synchrotron
emission due to relativistic particles, possibly (re)accelerated by shocks and gas turbulence, overlaid on the Bullet cluster X-ray image. }
\end{figure*}

\subsection*{How do hot baryons dynamically evolve in dark matter potentials?}

Clusters grow via accretion of dark and luminous matter along filaments
and the merger of smaller clusters and groups. X-ray observations show that
many present epoch clusters are indeed not relaxed systems, but are scarred
by shock fronts and contact discontinuities, and that the fraction of 
unrelaxed clusters likely increases with redshift. Although the gas 
evolves in concert with the dark matter potential, this gravitational assembly
process is complex, as illustrated by the temporary separations of dark and
X-ray luminous matter in massive merging clusters such as the ``Bullet 
Cluster'' (see Figure~1, left). In addition to the X-ray emitting hot gas,
the relativistic plasma seen through synchrotron emission in merging clusters
(Figure~1, right) is an important ICM component with at present
few observational constraints.

Major mergers are among the most energetic events in the Universe since the 
Big Bang,
releasing up to 10$^{64}$ ergs of gravitational potential energy through
the merger of two large subclusters. There are important questions to 
be answered, both to understand the complete story of galaxy and cluster
formation from first principles and, through a better understanding of
cluster physics, to increase the reliability of the constraints on
cosmological models derived from cluster observations (see white paper
by Vikhlinin et al.). These include: (1) How is the gravitational
energy that is released during cluster hierarchical formation dissipated in the
intracluster gas, thus heating the ICM, generating gas turbulence, 
and producing
significant bulk motions? (2) What is the origin and acceleration mechanism 
of the relativistic particles observed in the ICM? 
(3) What is the total level of nonthermal 
pressure support, which should be accounted for in the cluster 
mass measurements, and how does it evolve with time? To answer these questions, more than an order of magnitude improvement in spectral resolution is required,  while keeping good imaging capabilities, to map velocities and turbulence. \\

\begin{figure*}[tbh!]
\begin{minipage}[c]{1.0\textwidth}
\includegraphics[bb=88 266 528 722,scale=0.5]{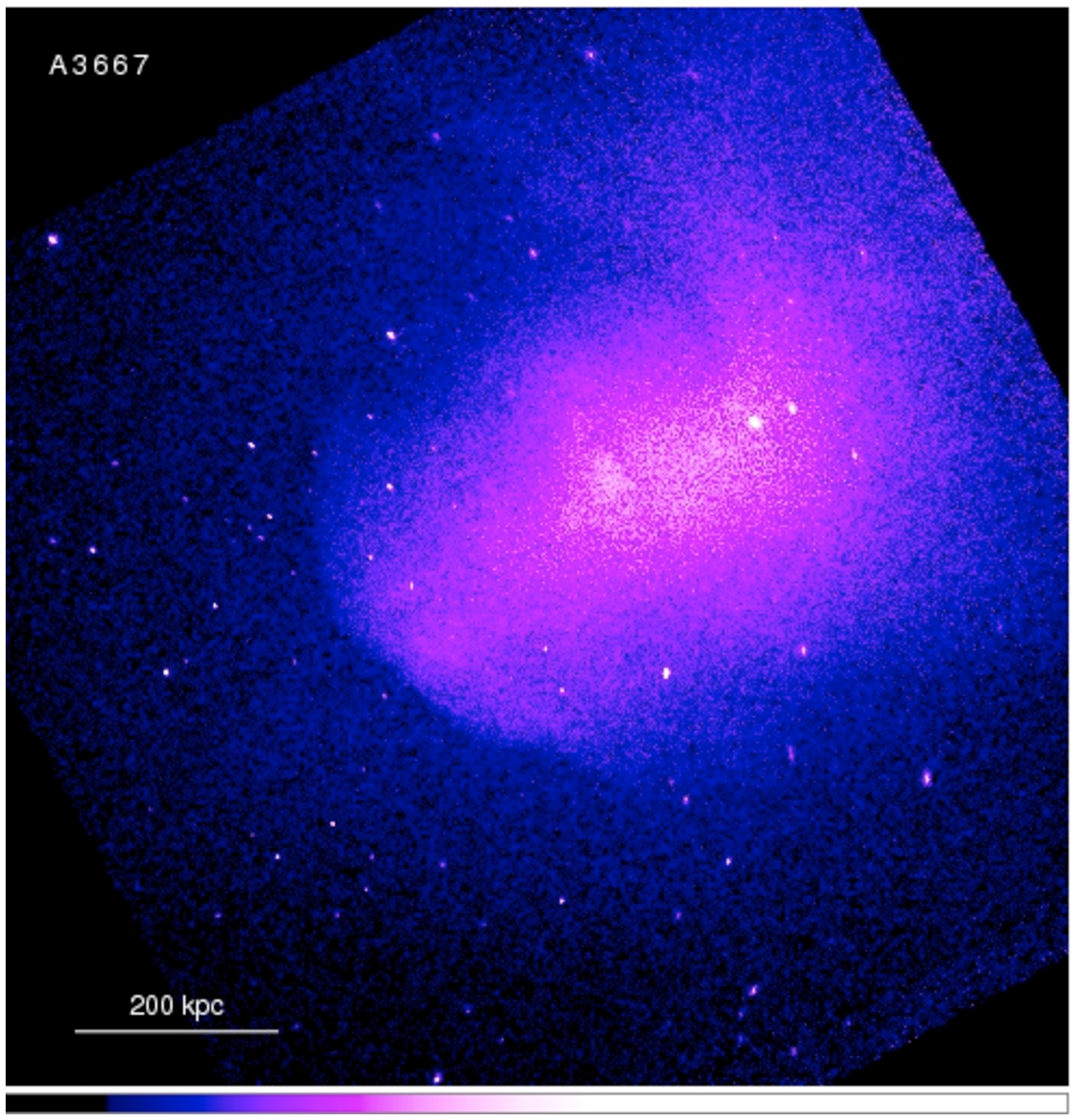}
\includegraphics[bb=82 273 507 701,scale=0.5]{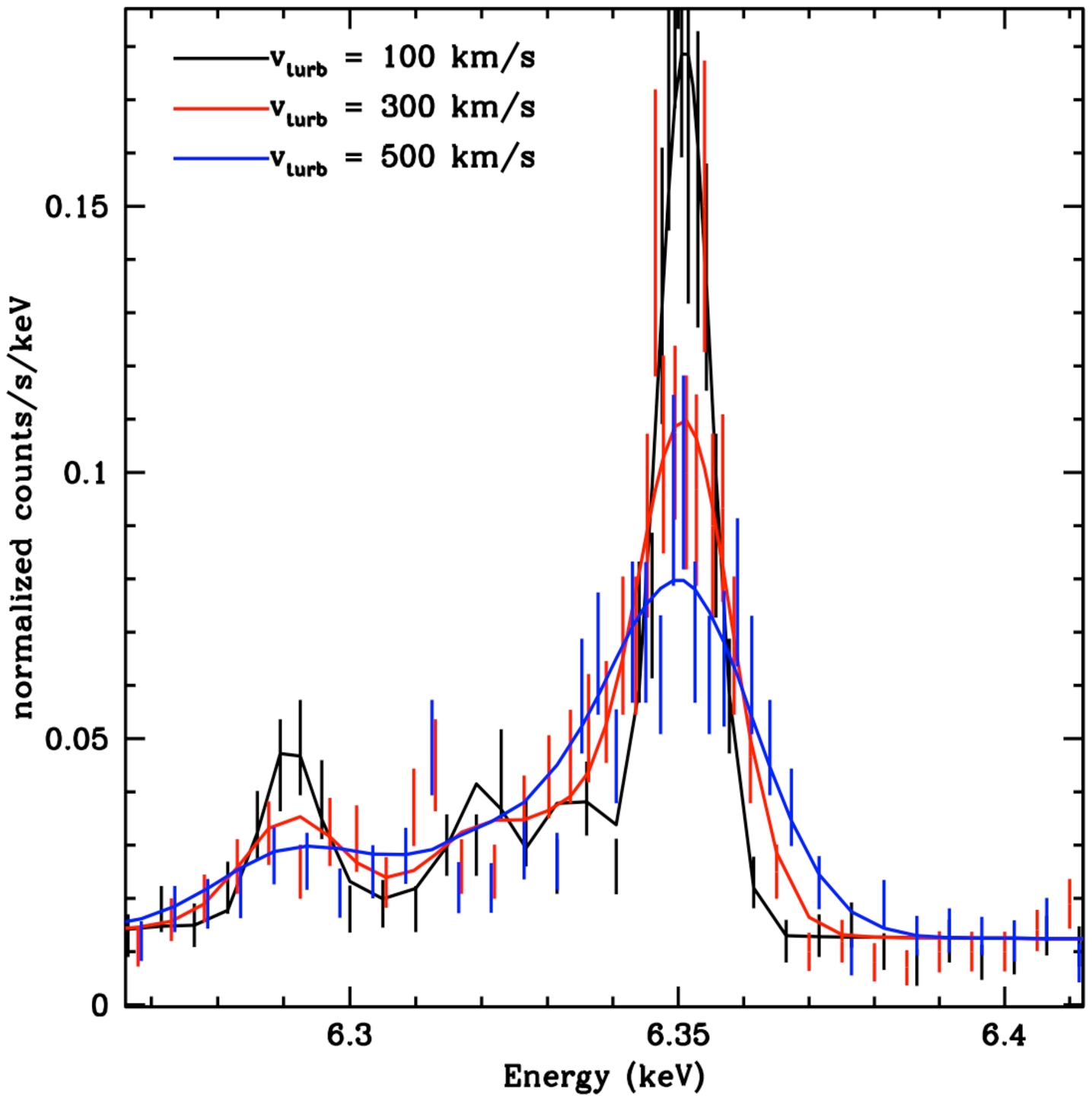}
\end{minipage}
\caption{{\bf Left:} 500 ksec {\em Chandra} image of the z=0.055 merging
cluster A3667. {\bf Right:} The 6 keV Fe line region 
of three simulated ({\em IXO}) calorimeter spectra for a 1 arcmin$^{2}$ region
in the very faint shock southwest of the A3667 merging subcluster. The 
exposure time of the simulations is 200 ksec. The three spectra correspond
to different levels of turbulence, with line widths of 100, 300, and
500 km s$^{-1}$. Since this cluster is undergoing a merger, we also expect
to see line shifts due to gas bulk motions.}
\end{figure*}

High-resolution X-ray spectral imaging can
determine the subcluster velocities and directions of motions by combining
redshifts measured from X-ray spectra (which give relative line-of-sight
velocities) and total subcluster velocities deduced from temperature and
density jumps across merger shocks or cold fronts [1]. 
These measurements combined with high quality 
lensing observations from instruments such as the LSST will probe how the hot gas
reacts in the evolving dark matter potential.

X-ray line width measurements will allow the level of gas turbulence to 
be mapped in detail for the first time.  As an example,
Figure~2  shows that the $2.5$ eV resolution of the  {\em IXO} calorimeter can distinguish line widths of
100, 300, and 500 km s$^{-1}$ in a  small (1 arcmin$^2$) region of the very faint shock
of the merging cluster A3667. For a high redshift cluster (with a luminosity 
of $\sim$ 10$^{44}$ erg s$^{-1}$ at z = 1), the
line width could still be measured to an accuracy of 70 km s$^{-1}$ in a
100 ksec exposure, and more precisely if more time is invested.

Sensitive hard X-ray (10--40 keV) imaging can reveal inverse Compton 
emission from the ICM. Although this emission has so far not even clearly
been detected, it promises unique information on the energy density of the
relativistic particles, and when combined with next generation
radio observatories like SKA, would probe the history of magnetic fields
in clusters. Capabilities like those of {\em IXO} are needed to understand
these observationally elusive, but important components of the ICM. \\


Further crucial insight into cluster assembly can be gleaned from measurements
of the {\it dark matter} distribution in the most relaxed systems. 
Cosmological numerical simulations of large-scale structure
collapse robustly predict that the dark matter distribution should be cuspy,
vary with system mass, and evolve with time. Current X-ray observations
of bright, local systems confirm the cuspy nature of the distribution, and
show some indication for a variation with mass [2,3]. 
With {\em IXO}, we can dramatically increase the
mass range available for these tests, and,  for the first time, tackle
the question of evolution  by determining
the mass profiles up to  high redshift ($z\sim 2$), even for low mass systems .

\subsection*{How and when was the excess energy in the intergalactic medium
generated?}

One of the most important revelations from X-ray observations, supported
by recent optical and IR studies, is that non-gravitational processes, 
particularly galaxy feedback from outflows created by 
supernovae and supermassive black holes 
(SMBH), must play a fundamental role, both in the history
of all massive galaxies and in the evolution of groups and clusters as a
whole. Galaxy feedback is likely to provide the extra energy required to keep
the gas in cluster cores from cooling all the way down to molecular
clouds, to account for the energy (i.e. entropy) excess observed in the
gas of groups and clusters, to cure the over-cooling problem, to regulate
star formation, and to produce the red sequence (Note the additional
discussion in the ``Cosmic Feedback'' white paper by Fabian et al.). 

It is now well established from {\em XMM-Newton} and {\em Chandra} observations
of local clusters and groups that their hot atmospheres have much more entropy
than expected from gravitational heating alone [4,5,6]. 
Determining when and how this 
non-gravitational excess energy was acquired will be an essential goal of the
next generation X-ray observatory. 
Galaxy feedback is a suspected source, but
understanding whether the energy was introduced early in the formation
of the first halos (with further consequence on galaxy formation history), or
gradually over time by AGN feedback, SN driven galactic winds, or an 
as-yet unknown physical process, is crucial to our understanding of how
the Universe evolved.

The various feedback processes, as well as cooling, affect the intergalactic
gas in different ways, both in terms of the level of energy modification
and the time-scale over which this occurs. Measuring the evolution of the gas
entropy and metallicity from the epoch of cluster formation is the key
information required to disentangle and understand the respective role for
each process. Since non-gravitational effects are most noticeable in groups
and poor clusters, which are 
the building blocks of today's massive clusters, these
systems are of particular interest. \\

A major challenging goal of a future next 
generation X-ray observatory is thus to study the properties of the first 
small clusters emerging at z$\sim$2 and directly trace their thermodynamic
and energetic evolution to the present epoch.

Future wide-field Sunyaev-Zel'dovich,  X-ray (e.g. SRG/eRosita) and optical-IR surveys will discover many thousands of clusters with z$<$2, but will provide
only limited information on their individual properties. These 
surveys will provide excellent samples of clusters for follow-up
IXO studies. In addition, 
$\sim$ 4 low mass clusters per deg$^{2}$, with 
M $>$ 10$^{13}$ M$_{\sun}$, will be detected serendipitously 
within the 18$\arcmin$ $\times$ 18$\arcmin$ field of the {\em IXO}
Wide Field Imager. Deep {\em IXO} observations will determine
the X-ray properties of even these low mass systems. 

The power of a high throughput, 
high resolution X-ray mission to study in detail high z 
clusters is illustrated in Figure~3 which shows simulated, deep spectra
for high redshift systems as would be obtained with the {\em IXO} calorimeter. These will provide gas density and temperature profiles, and thus entropy
and mass profiles to z $\sim$ 1 for low mass clusters (kT $\sim$ 2 keV, 
Fig.~3, middle) and for rarer more massive clusters, such as JKCS 041 
(Fig.~3, right) up to z $\sim$ 2, with a precision currently achieved only
for local systems. Measurements of the global thermal properties of the
first poor clusters in the essentially unexplored range z = 1.5--2 also
will become possible (Fig.~3, left). 

\begin{figure*}[tbh!]
\begin{minipage}[c]{1.0\textwidth}
\includegraphics[bb=90 344 500 705,scale=0.35]{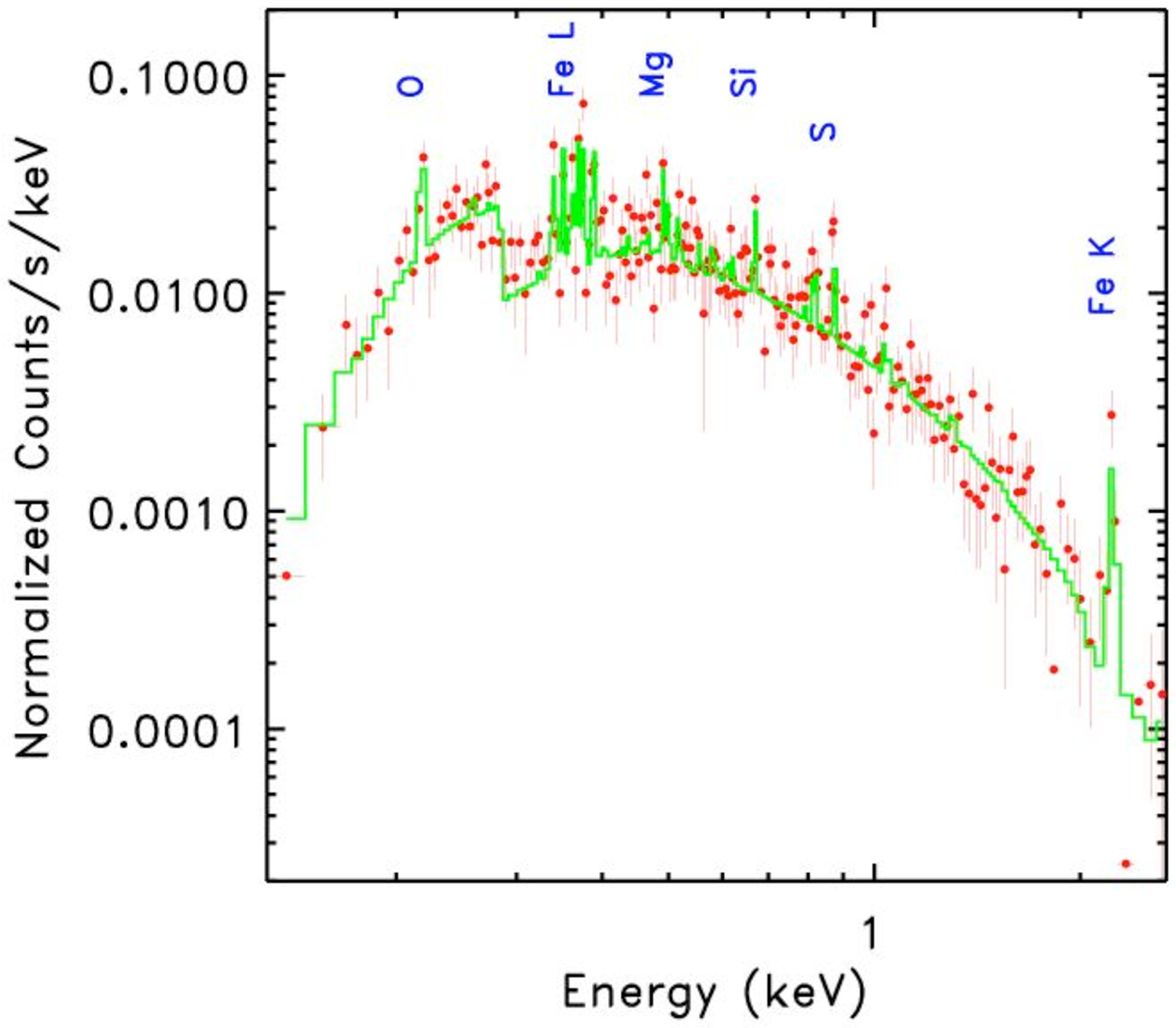}
\includegraphics[bb=84 352 513 709,scale=0.35]{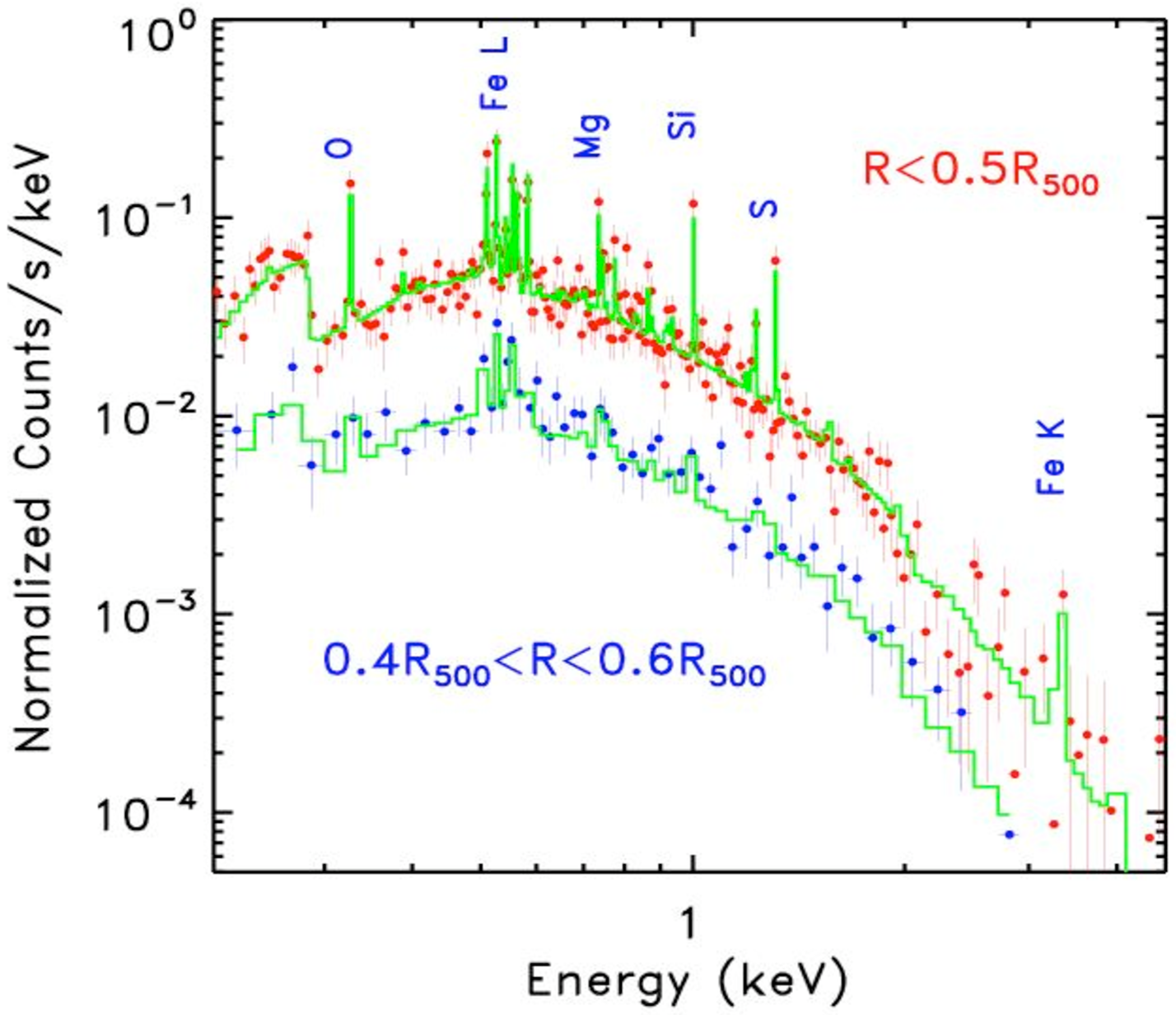}
\includegraphics[bb=91 268 525 701,scale=0.3]{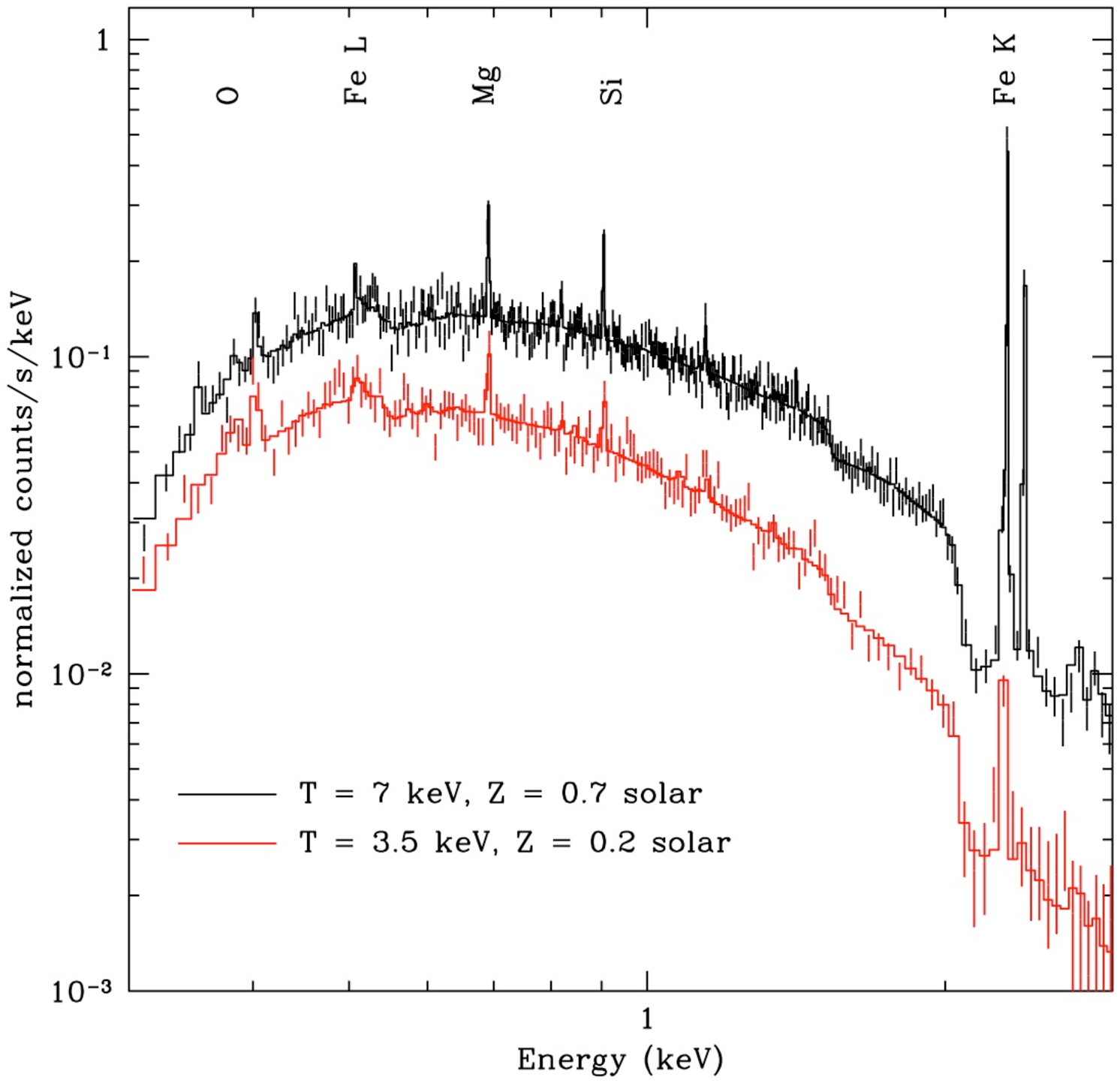}
\end{minipage}
\caption{X-ray spectra of high redshift clusters and groups will yield
gas temperatures and metallicity profiles. 
{\bf Left:} {\em IXO} 250 ksec observation of a low mass (kT=2 keV) 
z=2 cluster with a bolometric luminosity of 7.7 $\times$ 10$^{43}$ erg s$^{-1}$.
Overall temperature and abundances can be measured accurately: $\pm$3\%
for kT, $\pm$3.5\% for O and Mg, $\pm$25\% for Si and S, and $\pm$15\% for
Fe. {\bf Middle:} The same cluster, but at z=1, observed for 150 ks and
for two spectral extraction regions. In the 0.4R$_{500}$--0.6R$_{500}$ 
annulus (R$_{500}$ is a fiducial outer radius of the cluster where the
mean cluster mass density is a factor of 500 above the cosmic critical
density), the temperature and iron abundances are measured with an 
accuracy of $\pm$5\% and $\pm$20\% respectively, illustrating the 
capability of {\em IXO} to measure temperature and abundance profiles at z=1, 
even for low mass systems. {\bf Right:} Simulated {\em IXO} spectra
for the z=1.9 cluster JKCS 041 based on Chandra observations [7]. 
In an {\em IXO} exposure of 200 ks, the gas temperature of the core
(assumed to be 7 keV) is determined to 3\% uncertainty and the overall
metallicity to 3\%. In the outer region, the gas temperature (assumed to
be 3.5 keV) is equally well constrained, while the metallicity is measured
to 5\%.}
\end{figure*}

\subsection*{What is the cosmic history of heavy element production and
circulation?}

A fundamental astrophysical question is the cosmic history of heavy-element
production and circulation. This is strongly related to the history of
star formation, the time and environmental dependence of the stellar initial
mass function (and thus the cosmic history of Type I and II SNe) and 
the circulation of matter and energy between various phases of the Universe. 
As large 'closed' boxes of the Universe, clusters of galaxies are excellent
laboratories to study nucleosynthesis. While the next generation of 
optical/IR/sub-mm observatories such as JWST, TMT/GMT and ALMA will provide 
essential information on the star formation history, only sensitive X-ray measurements
of lines emitted by the hot ICM, where most of the metals reside, will 
directly determine the metal abundances in the ICM  to high redshifts.  \\

The first open question concerns the production of heavy elements over 
cosmic time. Chandra and XMM-Newton observations of clusters hint at Fe
abundance evolution from z=1 to the present [8,9].
To give a definitive answer on when the metals are produced,  we need 
to extend abundance studies to higher redshifts and for all astrophysically
abundant elements. In local systems, the abundance pattern of elements from O
to the Fe group, that are produced by supernovae, indicate that both type I
and type II SN contribute to the enrichment [10]. 
However, these measurements only provide a fossilized integral record of
the past SN enrichment and thus the evolution of supernovae remains largely
unconstrained. Furthermore, the main source of C and N, which can 
originate from a wide variety of sources (including stellar mass loss
from intermediate mass stars, whose cosmic history is poorly known) is
still under debate.

\begin{figure*}[tb]
\begin{minipage}[c]{1.0\textwidth}
\includegraphics[bb=89 350 543 722,scale=0.5]{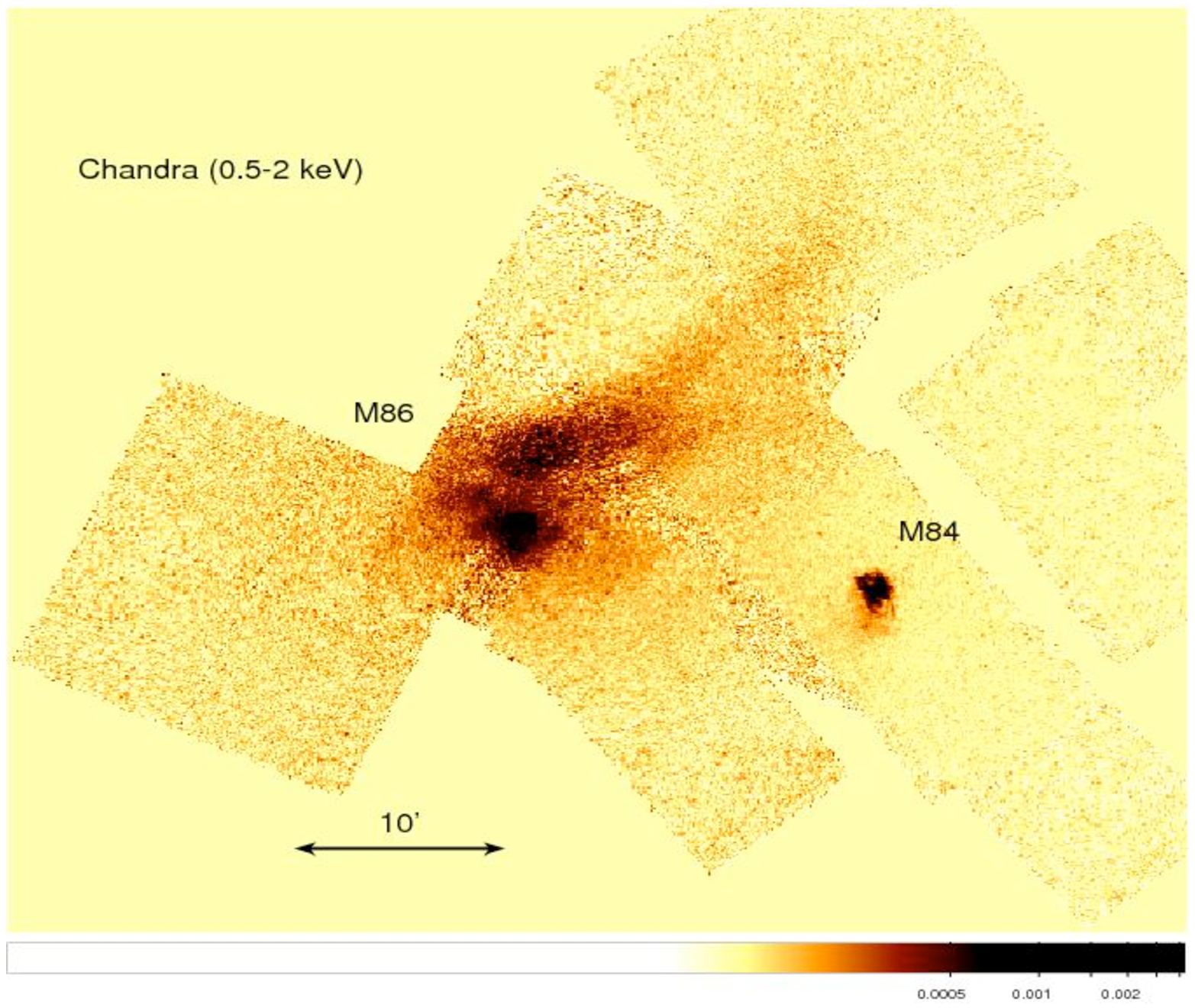}
\includegraphics[bb=89 232 552 723,scale=0.4]{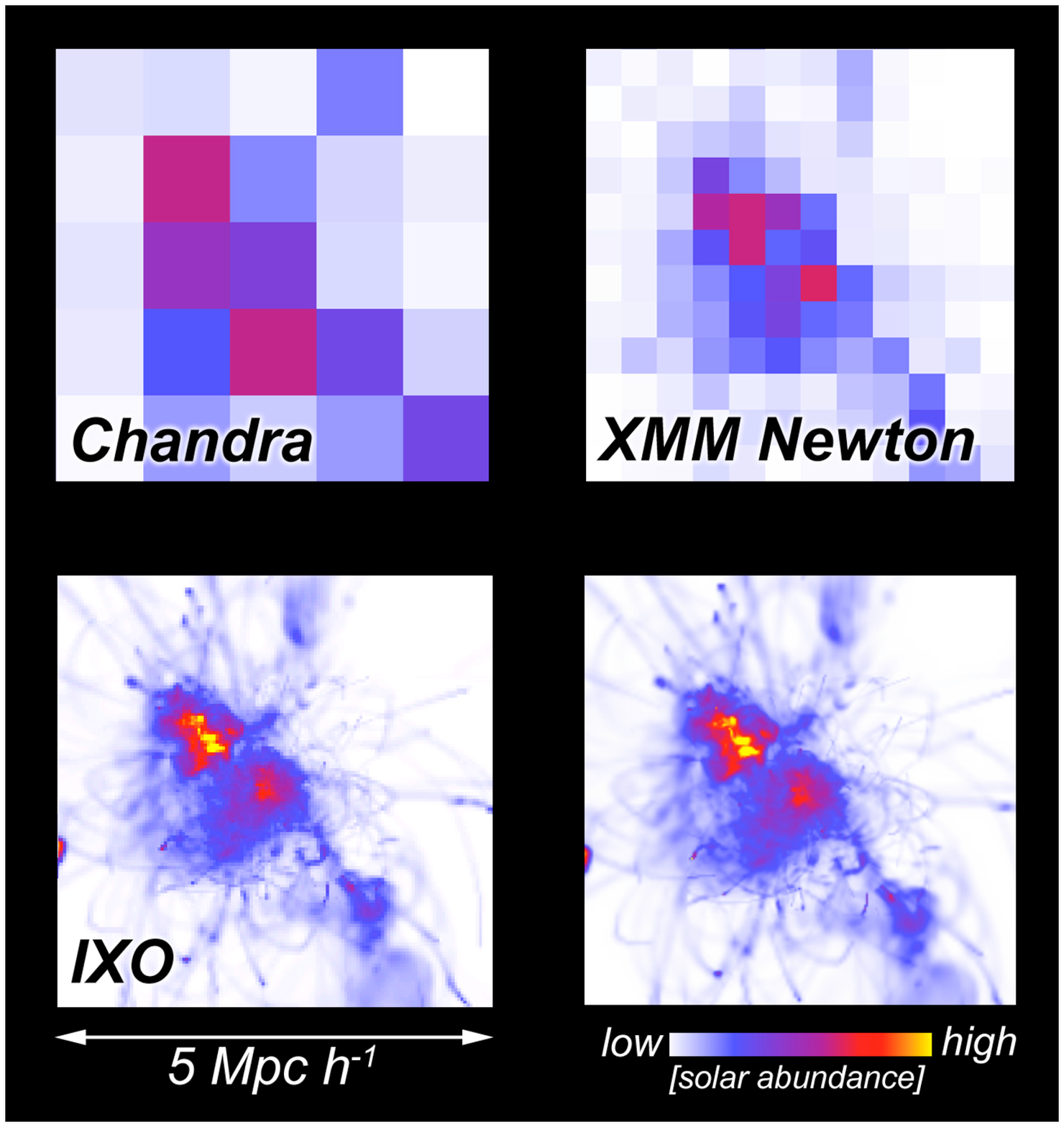}
\end{minipage}
\caption{{\bf Left:} X-ray emission from the Virgo elliptical galaxy 
M86 and its 380 kpc ram pressure stripped tail dominates this mosaic of 
Chandra images. Sensitive, high resolution X-ray spectroscopy 
will measure the metallicity along the tail and in the surrounding 
cluster gas. {\bf Right:} Chandra, 
XMM, and {\em IXO} 
simulations show the metal abundance and distribution in a 
merging cluster. In this example, a merging 7 keV cluster of 
L$_{\mathrm X}$ = 2 $\times$ 10$^{44}$ erg s$^{-1}$ at z = 0.05 is
observed for 30 ksec by each observatory compared to the actual assumed
metal distribution (lower right panel). In this example, the calorimeter
will image a 0.3$\times$0.3 Mpc region. A similar cluster at z=0.1 
would produce a comparable map in $\sim$ 100 ksec and the {\em IXO} 
field of view would be 0.54$\times$0.54 Mpc. Several exposures would
be required to map the whole region.}
\end{figure*}

The second fundamental -- and even more complex -- open question is how
the metals produced in the galaxies are ejected and redistributed in the
ICM. Although AGN outflows as well as galaxy--galaxy interactions can
add metals to the ICM, studies of local cluster abundance profiles suggest
that the metal enrichment of the ICM is due primarily to galactic winds
and ram pressure stripping of enriched gas from galaxies by the ICM
[10], an example being the ram-pressure stripped tail of the 
Virgo cluster galaxy M86, shown in Figure~4 (left). 
These processes would result in different 
distributions for the metallicity within the cluster and over time. For 
example, at high redshifts, galactic winds are expected to be effective
at enriching the ICM, while at lower redshifts, when massive clusters
have formed, the dense ICM, especially in the cluster cores, can ram-pressure
strip the enriched gas from galaxies and can even suppress galactic winds
and quench star formation. The enriched material is not expected to be
immediately mixed with the ICM, and in fact in the brightest, best studied
nearby clusters, current X-ray observations show that the metallicity 
distribution in the ICM is inhomogeneous [11].
 Thus by mapping the metallicity in samples of clusters over cosmic 
time, one
can untangle the various transport processes that contribute to the 
enrichment. In addition, if the mass in metals is calculated assuming that
the metallicity is uniform, this will miss estimate the true metal mass
in the clusters. 

Measurements of the metallicity distribution  in clusters, 
for a wide range of masses, dynamical states, and redshifts,  
are required in order to understand the ejection and redistribution 
process within clusters. These studies would also
have far reaching consequences for our understanding of the excess 
energy that is present in the ICM, as some of the transport processes
(SNe winds and AGN outflows) also inject energy into the ICM, as well as for
our understanding of environmental effects on the galaxy star formation
history, when combined with optical and IR observations. \\

A high-throughput, high-resolution X-ray observatory, such as {\em IXO}, 
is required
to provide answers to the questions concerning the production, circulation, 
and evolution of heavy elements in the ICM. The energy resolution of the
calorimeter, much smaller than the equivalent width of the strongest emission
lines, combined with good spatial resolution will allow a dramatic 
improvement in the abundance measurements. This is illustrated in Figures~3 and
4. Metal content and abundance patterns could be traced up to z $\sim$ 2 even 
in low mass clusters (Fig.~3, left). Element profiles will be measured up
to z $\sim$ 1 for poor clusters and up to z $\sim$ 2 for more massive objects
(Fig.~3, center and right). In addition, in more nearby clusters, we will
be able to resolve the 2D metal distribution down to the relevant physical
mixing scales (Fig.~4, right), and study in detail the process of 
metal injection by measuring the metallicity in the core and along the 
stripped tails of infalling galaxies (e.g. M86, Fig.~4, left). We will also be
able to measure for the first time, the abundance of trace elements like Mn to
Cr in a significant number of clusters. The production of these elements
by SNe Ia is very sensitive to the metallicity of the progenitor star and thus
X-ray spectroscopy will provide additional strong constraints on the cosmic
history of SNe enrichment.

\subsection*{Concluding Remarks}

Numerical simulations have reached a stage where modeling, including all
hydrodynamical and galaxy formation feedback processes, is becoming
feasible, although AGN feedback modeling is still in its infancy. The 
appropriate physics of these processes is not always clear, and advances
are largely driven by observation. Thus, constant
confrontation between numerical simulations of galaxy cluster formation and
observations is essential for making progress in the field. {\em IXO} 
observations
of the hot baryons, the most significant baryonic mass component of
clusters, combined with observations of the cold baryons
(from Herschel, JWST, ALMA, and ground based optical telescopes) as 
well as radio observations (e.g. SKA) will provide, for the first time, 
the details for a sufficiently critical comparison. We expect that the
major breakthrough of a detailed understanding of structure formation and 
evolution on cluster scales, as well as understanding 
the cosmic history of nucleosynthesis, 
will come from simulation--assisted interpretation and modeling of these
new generation observational data.

\subsection*{References}

{\bf 1:} Vikhlinin, A. et al.~2001, ApJ, 551, 160; {\bf 2:} Pointecouteau, E., 
et al.~2005, A\&A, 435, 1; {\bf 3:} Voigt, L., \& Fabian, A.~C., 2006, MNRAS, 
368, 618; {\bf 4:} Buote, D., et al.~2007, ApJ, 664, 123;
{\bf 5:} Pratt, G., et al.~2006, A\&A, 446, 429;  
{\bf 6:} Sun, M. et al.~2008, arXiv0805.2320; 
{\bf 7:} Andreon S., et al.~2008 arXiv0812.1699; 
{\bf 8:} Maughan, B., et al.~2008, 
ApJ.~Suppl., 2008, 174, 117; {\bf 9:} Balestra, I., et al.~2007,A\&A, 462, 429;
{\bf 10:} Kapferer, W., et al.~2007, A\&A, 466, 813; 
{\bf 11:} Sauvageot, J.-L., 2005, A\&A, 444, 673

\end{document}